\documentclass{elsart41}
\usepackage{graphics}
\usepackage{graphicx}
\usepackage{amssymb}
\begin{document}

\begin{frontmatter}



\title{PM-FM transition in a DE model}
%

\author[AA]{Eugene Kogan},
\ead{kogan@quantum.ph.biu.ac.il}
\author[BB]{Mark Auslender\corauthref{Mark Auslender}}
\ead{marka@ee.bgu.ac.il}

\address[AA]{Jack and Pearl Resnick Institute of Advanced Technology,
Department of Physics, Bar-Ilan University, Ramat-Gan 52900,
Israel}
\address[BB]{Department of Electrical and Computer Engineering,
Ben-Gurion University of the Negev,
P.O.B. 653, Beer-Sheva, 84105 Israel}

\corauth[Mark Auslender]{Corresponding author. Tel: +972 8 6461583
fax: +972 8 6472949}

\begin{abstract}

We study paramagnetic - ferromagnetic transition  due to
exchange interaction between  classical localized magnetic moments
and conduction electrons. By solving the Dynamical Mean
Field Approximation equations we  find  explicit formula for the transition
temperature $T_c$ for arbitrary  electron dispersion law, concentration  and  relation
between  exchange coupling and the electron band width.
We present the results of calculations of the $T_c$
for the semi-circular electron density of states.

\end{abstract}

\end{frontmatter}


The  double-exchange (DE) model \cite{zener,anderson,degennes} is one of the basic ones
in the theory of magnetism. Magnetic ordering appears in this model  due to exchange
coupling  between the core spins and the conduction electrons. The Hamiltonian  of the model
is
\begin{eqnarray}
\label{HamDXM}
H = \sum_{nn'\alpha} t_{n-n'} c_{n\alpha}^{\dagger} c_{n'\alpha}
-J \sum_{n\alpha\beta} {\bf S}_n\cdot
{\bf \sigma}_{\alpha\beta}c_{n\alpha}^{\dagger} c_{n\beta},
\end{eqnarray}
where $c$ and $c^{\dagger}$ are the electrons annihilation and creation
operators, ${\bf S}_n$ is the operator of a core spin,  $t_{n-n'}$ is the
electron hopping,
$J$ is  the
exchange
coupling between a core spin and n electrons,
$\hat{\bf \sigma}$ is the vector of the Pauli matrices, and $\alpha,\beta$ are
spin indices.

We calculate the temperature  of a paramagnetic-ferromagnetic
transition $T_c$ in a double-exchange model  for arbitrary
electron dispersion law, concentration  and  relation
between the exchange coupling and the electron band width by formulating and
solving the DMFA equations. We treat the core spins as classical vectors.
The DE Hamiltonian in a single electron representation
can be presented as
\begin{equation}
\label{generic}
H_{nn'}=t_{n-n'}-J {\bf m}_n\cdot {\bf \sigma}\delta_{nn'}.
\end{equation}
Let us introduce Green's function and local Green's function
\begin{eqnarray}
\label{green}
\hat{G}(E)=(E-H)^{-1},\qquad
\hat{G}_{\rm loc}(E)=
\left\langle\hat{G}_{nn}(E)\right\rangle.
\end{eqnarray}
In the last equation, the averaging is with respect to random configurations of
the core spins.
In the framework of the DMFA approach to the problem (see \cite{DMFA}
and references therein) the local Green's function
is expressed through the  the local self-energy $\hat{\Sigma}$ by the
equation
\begin{eqnarray}
\label{local}
\hat{G}_{\rm loc}(E) =g_0\left(E - \hat{\Sigma}(E)\right),
\end{eqnarray}
where $g_0(E) =\frac{1}{N}\sum_{\bf k}\left(E-t_{\bf k}\right)^{-1}$
is the bare (in the
absence of the  exchange interaction) local Green's function. The
self-energy satisfies equation
\begin{eqnarray}
\hat{G}_{\rm loc}(E)=\left\langle \frac{1}
{\hat{G}_{\rm loc}^{-1}(E)+\hat{\Sigma} (E)
+J{\bf m}\cdot\hat{\bf \sigma}}\right\rangle,
\label{cpa}
\end{eqnarray}
where $\left\langle X({\bf m})\right\rangle \equiv \int X({\bf m})P({\bf m})$,
and $P({\bf m})$ is a probability
of a given spin orientation (one-site probability). The quantities
$\hat{G}$ and $\hat{\Sigma}$
are $2\times 2$ matrices in spin space.

The DMFA assumption for the probability $P({\bf m})$ is based on the Equation
\begin{eqnarray}
\label{probability4}
\Delta D(E,{\bf m})= -\frac{1}{\pi}{\rm Im}
\ln {\rm det}\left[1+\left(J{\bf m}\hat{\bf \sigma}
+\hat{\Sigma}\right)\hat{G}_{\rm loc}\right],
\end{eqnarray}
where the argument of both $G_{\rm loc}$ and $\Sigma$ is $E+i0$.
So the change in thermodynamic potential is  \cite{doniach,chat,ak2}
\begin{eqnarray}
\Delta\Omega({\bf m})=\int f(E)\Delta D(E,{\bf m})dE,
\label{probability}
\end{eqnarray}
where  $f(E)$ is the Fermi function.
The DMFA approximation  for the one-site
probability $P({\bf m})$ is:
\begin{eqnarray}
\label{prob2}
P({\bf m})\propto \exp\left[-\beta\Delta\Omega({\bf m})\right].
\end{eqnarray}

Eqs. (\ref{cpa}) and (\ref{prob2}) are the system of  non-linear (integral)
equations. In the ferromagnetic (FM) phase near the
Curie temperature, Eqs. (\ref{cpa}) and (\ref{prob2}) can be linearized
\cite{chat} with respect to
${\bf M}$. Thus
we  reduce the DMFA equations
to a traditional MF equation  \cite{ak2}
\begin{eqnarray}
P({\bf m})\propto  \exp\left( -3\beta T_c{\bf M}\cdot{\bf m}\right).
\label{probability2}
\end{eqnarray}
The parameter $T_c$ is formally introduced as a coefficient in the
expansion of $\Delta \Omega({\bf m})$ with respect to
${\bf M}$.
Non-trivial solution of the MF equation
${\bf M}=\left\langle{\bf m}\right\rangle$
can exist only for $T<T_{c}$, hence $T_c$ is the ferromagnetic transition
temperature.

For the $T_c$, after straightforward, though lengthy algebra, we
obtain
\begin{eqnarray}
\label{Theta}
T_{c}=\frac{2J^2}{3\pi}\int_{-\infty}^{\mu}
\mbox{Im}\left[\frac{g}{
\frac{(\Sigma g'-g)(1+\Sigma g)}
{g'+g^2} -\frac{2J^2g}{3}}\right]dE,
\end{eqnarray}
where $\Sigma$ and $g$ are determined by the properties of the
system in the PM phase ($P({\bf m})={\rm const}$). Eq.
(\ref{Theta}) is  the main result of our paper. Let us apply Eq.
(\ref{Theta}) to the case of semi-circular bare density of states
$N_0(\varepsilon)=(2/W)\sqrt{E^2/W^{2}-1}$, for which Eq.
(\ref{Theta}) takes the form
\begin{eqnarray}
\label{Thetasc}
T_{c}= \mbox{Im}\int_{-\infty}^{\mu}
\frac{\frac{J^2W^2}{6\pi}g^2dE}
{\left(E-\frac{W^2g}{4}\right)\left(E-\frac{W^2g}{2}\right) -
\frac{J^2W^2g^2}{6}}.
\end{eqnarray}

For arbitrary exchange, the integral in Eq. (\ref{Thetasc}) can be
calculated only numerically.
\begin{figure}[!ht]
\begin{center}
\includegraphics[width=0.45\textwidth]{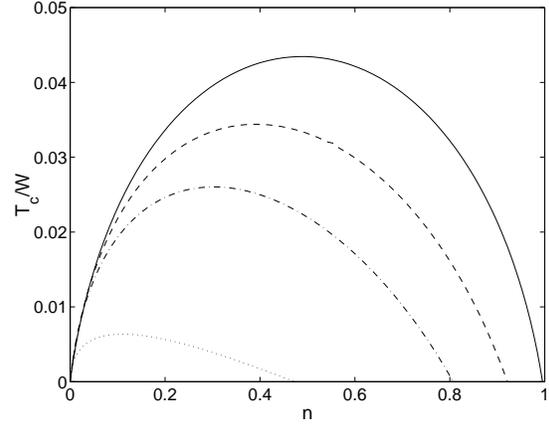}
\end{center}
\caption{$T_c$ as a function of electron concentration $n$:
  $J/W=.25$ (dotted line), $J/W=1$ (dash-dotted
 line),  $J/W=2$ (dashed line), and $J/W=20$ (solid line)}
\label{FIG1}
\end{figure}
Note that  that our main result (equation for the $T_c$) indicates
it's own limits of validity. In part of the $J/W-n$ plane,  Eq.
(\ref{Thetasc}) gives $T_c<0$. Negative value of $T_c$ means, that
at any temperature, including $T=0$,  the paramagnetic phase is
stable with respect to  appearance of small spontaneous magnetic
moment,  and strongly suggests that the ground state in this part
of the phase plane is nonferromagnetic (NFM) \cite{chat3}.

\begin{figure}[!ht]
\begin{center}
\includegraphics[width=0.45\textwidth]{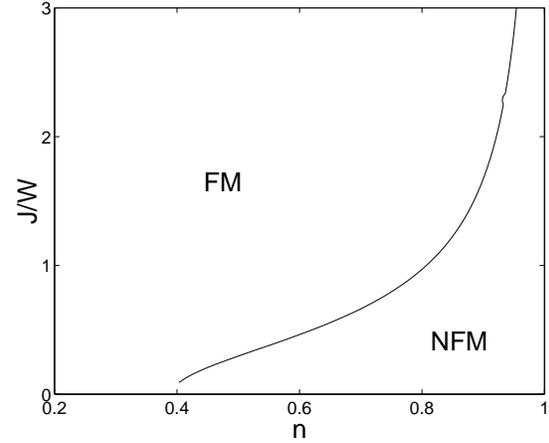}
\end{center}
\caption{The curve $T_c=0$  on the $J/W-n$ plane}
\label{FIG2}
\end{figure}


\begin{thebibliography}{99}

\bibitem{zener} C. Zener, Phys. Rev. {\bf 82}, 403 (1951).

\bibitem{anderson} P. W. Anderson and H. Hasegawa, Phys. Rev. {\bf 100}, 675
(1955).


\bibitem{degennes} P. G. De Gennes, Phys. Rev. {\bf 118}, 141 (1960).






\bibitem{DMFA} A. Georges, G. Kotliar, W. Krauth, and M. J. Rozenberg, Rev. Mod.
Phys. {\bf 68}, 13 (1996).



and S. Mahanti (Plenum Publishing, New York, 1999).


\bibitem{chat}  A. Chattopadhyay and A. J. Millis
Phys. Rev. B {\bf 64}, 024424 (2001);
A. Chattopadhyay, S. Das Sarma, and A. J. Millis
Phys. Rev. Lett. {\bf 87}, 227202 (2001).


\bibitem{doniach} S. Doniach and E. H. Sondheimer, {\it Green's functions for
solid
state physicists} (Imperial College Press, London, 1998).


\bibitem{ak2}E. Kogan and M. Auslender, \hfill \break
Phys. Rev. B {\bf 67}, 132410 (2003).


\bibitem{chat3}
A. Chattopadhyay, A. J. Millis, and S. Das Sarma
Phys. Rev. B {\bf 64}, 012416 (2001).


\end{thebibliography}
\end{document}